\documentclass[review]{elsarticle}

\usepackage{lineno,hyperref}
\modulolinenumbers[5]

\journal{Photon. Nanostruct. Fundam. Appl.}

\usepackage{color}

\begin{document}

\begin{frontmatter}

\title{Analogue transformation acoustics and the compression of spacetime}


\author[NTC]{Carlos Garc\'{i}a Meca\corref{mycorrespondingauthor}}
\ead{cargarm2@ntc.upv.es}
\author[PCU]{Sante Carloni}
\author[IAA]{Carlos Barcel\'{o}}
\author[UC3]{Gil Jannes}
\author[WPG]{Jos\'{e} S\'{a}nchez-Dehesa}
\author[NTC]{Alejandro Mart\'{i}nez}

\cortext[mycorrespondingauthor]{Corresponding author}

\address[NTC]{Nanophotonics Technology Center, Universitat Polit\`{e}cnica de Val\`{e}ncia, Camino de Vera s/n, 46022 Valencia, Spain}
\address[PCU]{Institute of Theoretical Physics, Faculty of Mathematics and Physics, Charles University in Prague, 
V. Hole\v{s}ovi\v{c}k\'{a}ch 2 180 00 Praha 8,  Czech Republic}
\address[IAA]{Instituto de Astrof\'{i}sica de Andaluc\'{i}a (CSIC), Glorieta de la Astronom\'{i}a, 18008 Granada, Spain}
\address[UC3]{Modelling \& Numerical Simulation Group, Universidad Carlos III de Madrid,
Avda. de la Universidad, 30, 28911 Legan\'{e}s (Madrid), Spain}
\address[WPG]{Wave Phenomena Group, Universitat Polit\`{e}cnica de Val\`{e}ncia, Camino de Vera s/n, 46022 Valencia, Spain}


\begin{abstract}
A recently developed technique known as analogue transformation acoustics has allowed the extension of the transformational paradigm to general spacetime transformations under which the acoustic equations are not form invariant. In this paper, we review the fundamentals of analogue transformation acoustics and show how this technique can be applied to build a device that increases the density of events within a given spacetime region by simultaneously compressing space and time.
\end{abstract}


\end{frontmatter}

\linenumbers

\section{Introduction}
Metamaterials offer an unprecedented flexibility in the construction of media with properties that are difficult or impossible to find in nature~\cite{ZHA10-CSR}. This concept first appeared within the frame of electromagnetism, and enabled scientists to design exotic devices such as negative-index superlenses~\cite{PEN00-PRL}. Afterwards, the notion of metamaterial has been extended to other branches of physics~\cite{WEG13-SCI}, such as acoustics~\cite{NOR09-JASA,TOR09-PRB}, electronics~\cite{SIL12-PRB} or thermodynamics~\cite{GUE12-OE,SCH13-PRL}. To take full advantage of this flexibility in the synthesis of tailor-made properties, we also need new design techniques that help us to engineer these properties with the aim of building novel devices with advanced functionalities. Along this line, one of the most powerful techniques is transformation optics, which prescribes the properties that a medium should have in order to alter the propagation of light in almost any imaginable way~\cite{LEO06-SCI,PEN06-SCI,SHA08-SCI,CHE10-NM}. As a result, metamaterials and transformation optics have teamed up to open the door to the realization of photonic devices that were unthinkable only a few years ago, such as invisibility cloaks or optical wormholes~\cite{SCH06-SCI,GRE07-PRL}, constituting one of the most interesting recent developments in material science. The great success of the transformational paradigm in the field of electromagnetism has led the research community to look for ways in which this approach could be extended to other fields~\cite{ZHA08-PRL,CHE10-JPD,GUE12-OE}. 

Noticing that the key to transformation optics is the form invariance of Maxwell's equations under any spacetime transformation, the initial approach was to try to exploit form invariance in the governing equations of different physical phenomena. Therefore, one of the crucial issues in transformational methods is the range of coordinate transformations over which the relevant field equations have this property~\cite{Milton,NOR11-WM,BER13-ARX,GAR14}. Outside of optics, acoustics is probably the field in which the greatest advance has been achieved. There, the form invariance of the acoustic equations under spatial transformations has been used to obtain the material parameters that deform acoustic space in the desired way, e.g., for cloaking acoustic waves~\cite{NOR09-JASA, CHE10-JPD, CUM07-NJP, CHE07-APL, TOR08-NJP, PEN08-NJP, FAR08-NJP, POP11-PRL, CRASTER}. 

However,  this approach to transformation acoustics has been undermined by the deep structural differences between Maxwell's theory with its underlying relativistic geometry on the one hand, and the Galilean character of fluid mechanics on the other hand, which reduces the power of the traditional transformational method when applied to acoustics. Specifically, classical acoustic equations are not form invariant under transformations that mix space and time~\cite{GAR14}. As a consequence, the method cannot be applied to design devices based on this kind of transformation, contrarily to what has been done in optics~\cite{MCC11-JOPT,FRI12-NAT,CUM11-JOPT}. 

Recently, the problem of transformation acoustics was approached from another angle~\cite{GAR13-SR,GAR14}. Instead of using directly the symmetries of the acoustic equations to bridge between different solutions for the propagation of acoustic waves, the symmetries of an analogue abstract spacetime (described by relativistic form-invariant equations) are exploited. In this method, each couple of solutions connected by a general coordinate transformation in the analogue spacetime can be mapped to acoustic space. This way, it is possible to find the relation between the acoustic material parameters associated with each of these transformation-connected solutions. The result is an alternative version of transformation acoustics as powerful as its optical counterpart and that we refer to as analogue transformations acoustics (ATA).

In this paper, we review the ATA method and some of the devices it has allowed us to engineer, which were unworkable through other approaches (section~\ref{ATA}). Subsequently, we use this technique to design a new kind of spacetime compressor that increases the density of events in a given region. The performance of the device is analyzed and verified through numerical calculations (section~\ref{Compressor}). The differences between the proposed spacetime compressor and other squeezing devices are discussed. Finally, some conclusions are drawn~\ref{Conclusion}.

\section{Analogue transformation acoustics} \label{ATA}
The extension of transformation acoustics to general spacetime transformations presents two separate problems. On the one hand, the acoustic equations are not form invariant under transformations that mix space and time. As mentioned above, this drawback can be circumvented with the aid of an auxiliary relativistic spacetime. On the other hand, it has been shown that the acoustic systems usually considered in transformation acoustics (which deal with the propagation of acoustic waves in stationary or non-moving fluids) do not posses enough degrees of freedom so as to mimic an arbitrary spacetime transformation. This is the case of the system represented by the standard pressure wave equation~\cite{GAR14}. The limitations of transformational pressure acoustics have also been analyzed by other authors~\cite{KIN14-ARX}. 

Instead of the pressure wave equation, ATA uses the wave equation for the velocity potential $\phi_1$ (defined as $-\nabla \phi_1= \mathbf{v_1}$, where $\mathbf{v}_1$ is the velocity of the acoustic perturbation), which reads
\begin{equation}
\label{wave_moving}
-\partial_t\left(\rho_{\rm V}{c_{\rm V}}^{-2}\left(\partial_t\phi_1 +\mathbf{v}_{\rm V}\cdot\nabla\phi_1\right)\right)+\nabla\cdot\left(\rho_{\rm V}\nabla\phi_1-\rho_{\rm V} c_{\rm V}^{-2}\left(\partial_t\phi_1 + \mathbf{v}_{\rm V}\cdot\nabla\phi_1 \right)\mathbf{v}_{\rm V}\right)=0,
\end{equation}
where $\mathbf{v}_{\rm V}$ (background velocity), $\rho_{\rm V}$ (mass density) and $c_{\rm V}$ (speed of sound) will be considered as the acoustic properties of virtual space. 
There are two reasons behind the choice of Eq.~\ref{wave_moving}. First, although it is not form invariant under general spacetime transformations either, there is a well-known relativistic model that is analogue to this equation~\cite{VIS98-CQG, BAR11-LRR}. Second, Eq.~\ref{wave_moving} allows us to consider the propagation of waves in a moving fluid, which provides the missing degrees of freedom. 

If we directly applied a coordinate transformation mixing space and time to Eq.~\ref{wave_moving}, there would appear new terms that could not be ascribed to any property of the medium. The ATA method starts by momentarily interpreting this equation as a different one with better transformation properties. In particular, we use the fact that Eq.~\ref{wave_moving} can be written as the massless Klein-Gordon equation of a scalar field $\phi_1$ propagating in a (3+1)-dimensional pseudo--Riemannian manifold (the abstract spacetime)~\cite{VIS98-CQG, BAR11-LRR}: 
\begin{equation} \label{dAlembertian}
{1\over\sqrt{-g}} \partial_\mu \left( \sqrt{-g} \; g^{\mu\nu} \; \partial_\nu \phi_1 \right)=0,
\end{equation}
where $g_{\mu\nu}$ is the 4-dimensional metric (with $g$ its determinant) of the abstract spacetime. The inverse metric $g^{\mu\nu}$ is given by
\begin{eqnarray}
g^{\mu\nu} = \frac{1}{\rho_{\rm V} c_{\rm V}}
\left(
\begin{array}{ccc}
-1 & \vdots & -v_{\rm V}^i \\
... & . & ................ \\
-v_{\rm V}^i & \vdots & c_{\rm V}^2\tilde{\gamma}^{ij}-v_{\rm V}^iv_{\rm V}^j
\end{array}
\right)~.
\end{eqnarray}
We must stress that, although Eqs.~\ref{wave_moving} and \ref{dAlembertian} are formally identical when expressed in, for instance, a Cartesian coordinate system, they are completely different equations due to the contrasting nature of the elements that appear in each one. While $\rho$ and $c$ are scalars and $\mathbf{v}$ is a three-dimensional vector, $g_{\mu\nu}$ is a four-dimensional tensor. Thus, each equation behaves differently under general transformations. The point in interpreting momentarily Eq.~\ref{wave_moving} as Eq.~\ref{dAlembertian} is that the latter preserves its form under any transformation, including the ones that mix space and time.

The next step is to apply the desired transformation $\bar{x}^{\mu}=f(x^{\mu})$ to Eq.~\ref{dAlembertian}. This just implies replacing the metric $g^{\mu\nu}$ by another one $\bar{g}^{\mu\nu}$, which can be derived from the former by using standard tensorial transformation rules 
\begin{equation} \label{metric_transformation}
\bar{g}^{\bar{\mu}\bar{\nu}}=\Lambda^{\bar{\mu}}_{\mu}\Lambda^{\bar{\nu}}_{\nu}g^{\mu\nu},
\end{equation}
where $\Lambda^{\bar{\mu}}_{\mu}=\partial{\bar{x}^{\bar{\mu}}}/\partial{x^{\mu}}$.
This way we obtain the transformed version of Eq.~\ref{dAlembertian}
\begin{equation} \label{dAlembertian2}
{1\over\sqrt{-\bar{g}}} \partial_\mu \left( \sqrt{-\bar{g}} \; \bar{g}^{\mu\nu} \; \partial_\nu \bar{\phi}_1 \right)=0.
\end{equation}
Finally, if we relabel the coordinates $\bar{x}^\mu$ of Eq.~\ref{dAlembertian2} to $x$, we can interpret Eq.~\ref{dAlembertian2} as the velocity potential wave equation (expressed in the same coordinate system as Eq.~\ref{wave_moving}, e.g., Cartesian) associated with a medium characterized by acoustic parameters $\mathbf{v}_{\rm R}$, $\rho_{\rm R}$ and $c_{\rm R}$
\begin{equation}
\label{wave_moving2}
-\partial_t\left(\rho_{\rm R}{c_{\rm R}}^{-2}\left(\partial_t\bar{\phi}_1 +\mathbf{v}_{\rm R}\cdot\nabla\bar{\phi}_1\right)\right)+\nabla\cdot\left(\rho_{\rm R}\nabla\bar{\phi}_1-\rho_{\rm R} c_{\rm R}^{-2}\left(\partial_t\bar{\phi}_1 + \mathbf{v}_{\rm R}\cdot\nabla\bar{\phi}_1 \right)\mathbf{v}_{\rm R}\right)=0,
\end{equation}
which will represent physical space. That is, we identify
\begin{eqnarray}
\bar{g}^{\mu\nu} = \frac{1}{\rho_{\rm R} c_{\rm R}}
\left(
\begin{array}{ccc}
-1 & \vdots & -v_{\rm R}^i \\
... & . & ................ \\
-v_{\rm R}^i & \vdots & c_{\rm R}^2\tilde{\gamma}^{ij}-v_{\rm R}^iv_{\rm R}^j
\end{array}
\right)~.
\end{eqnarray}
This last equality establishes a relation between the acoustic parameters of virtual and physical space, as $\bar{g}^{\mu\nu}$ is also a function of $\mathbf{v}_{\rm V}$, $\rho_{\rm V}$ and $c_{\rm V}$ (remember that it comes from $g^{\mu\nu})$. As mentioned above, Eqs.~\ref{wave_moving} and \ref{dAlembertian} are analogous (formally, they have the same mathematical solutions), and the same happens with Eqs.~\ref{dAlembertian2} and \ref{wave_moving2}. These connections guarantee that the velocity potentials $\phi_1$ and $\bar{\phi}_1$ in virtual and physical spaces are related by the applied transformation, since so are the fields $\phi_1$ and $\bar{\phi}_1$ that appear in Eqs.~\ref{dAlembertian} and \ref{dAlembertian2}. 

Following this procedure, we can now work with transformations that mix space and time. As an example, ATA has been used to design the acoustic counterparts of an electromagnetic frequency converter~\cite{CUM11-JOPT} and a time cloak~\cite{MCC11-JOPT}, both based on spacetime transformations, which was not possible with the standard method based on direct transformations of the acoustic equations~\cite{GAR14,GAR13-SR}.

\begin{figure}[]
\begin{center}
  \includegraphics{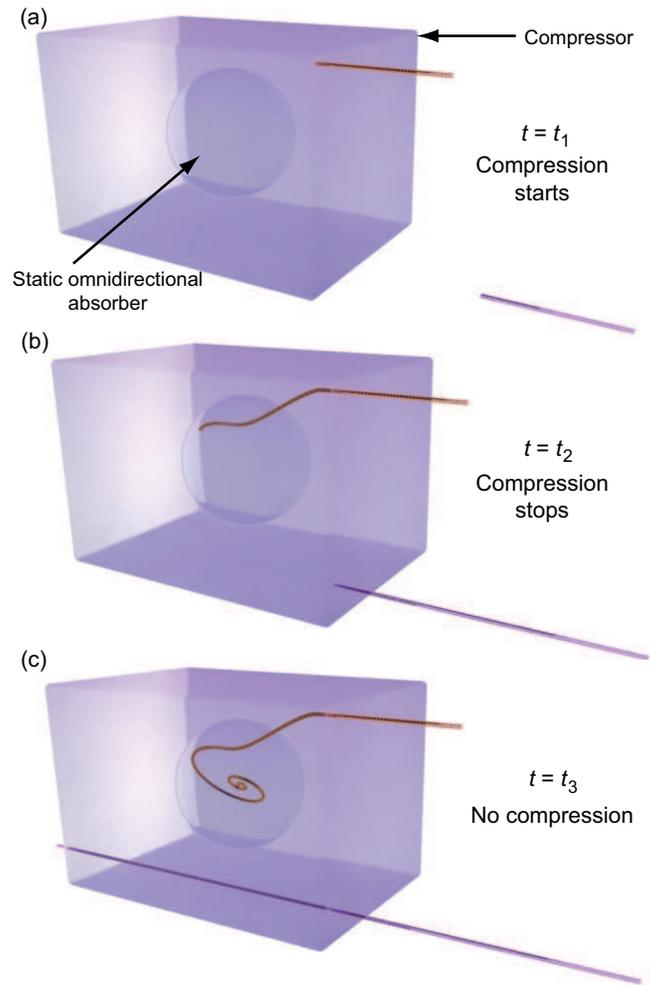}
\end{center}
  \caption{Compressor-absorber system. The simulated trajectories of two acoustic rays that enter the compressor at different instants are shown.
The absorber has a spherical shape with radius $R$, and is characterized by a static refractive index (relative to the background medium surrounding the compressor) given by $n=0.5 R/r$, where $r$ is the distance from its center~\cite{GAR13-SR,CLI12-APL}.}
  \label{Fig1}
\end{figure}

Another more elaborate example is that of a dynamically tunable space compressor given by the transformation $\bar{t}=t$ and $\bar{x}^i=x^if_0(t)$, which acts only inside a three-dimensional box (the compressor) [note that we use Latin spatial indices ($i,j$) and Greek spacetime indices ($\mu, \nu$, with $x^0=t$) ]. That is, space is uniformly compressed as a function of time inside this box (this is controlled by the value of $f_0$), while the time variable remains unchanged. As an application of this gadget, it was proposed to combine the compressor with an omnidirectional absorber placed inside it, in order to select which rays are trapped by the absorber~\cite{GAR13-SR}. The simulated performance of a specific configuration of the compressor-absorber device is shown in Fig.~\ref{Fig1}. In this case, $f_0(t)$ was chosen such that compression starts smoothly at $t = t_1$ [$f_0(t<t_1)=1$], just after the orange ray enters the box. Compression forces this ray to approach the absorber, which ultimately traps it. Later on, at $t=t_2$, compression fades away [$f_0(t>t_2)=1$]. The purple ray enters the box soon after and thus follows a straight line.  

%
\section{Compressing spacetime} \label{Compressor}
\begin{figure}[]
\begin{center}
  \includegraphics{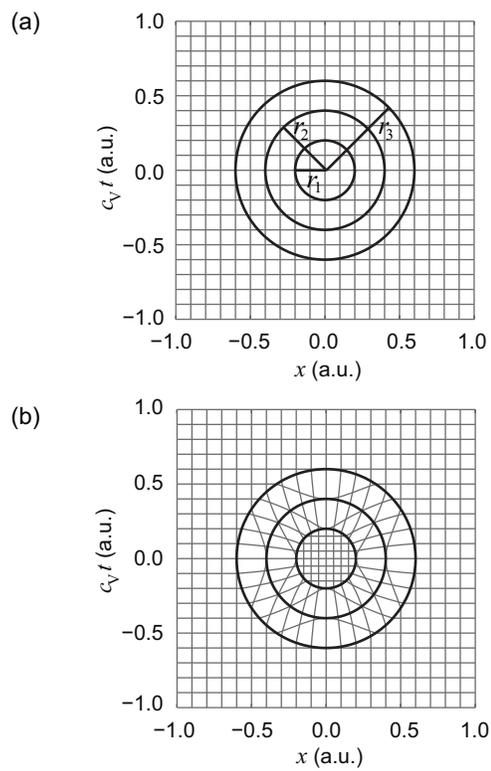}
\end{center}
  \caption{Transformation that simultaneously compresses space and time. (a) Cartesian grid in virtual space. (b) Grid deformed by the proposed transformation.}
  \label{Fig2}
\end{figure}

In the previous section we described a device that dynamically compresses space without changing the time variable. In practice, this compression can only be performed inside a box. As a consequence, we create a discontinuity at the box boundaries that can produce reflections. To avoid these reflections, we must ensure that there is no compression at the instant rays enter or exit the box. In this section, we analyze the possibility of performing a continuous transformation of both space and time that eliminates this problem. Specifically, we will consider the mapping ($\bar{x}=x\bar{r}/r$ ; $\bar{t}=t\bar{r}/r$), with
\begin{equation}
r=\sqrt{x^2+(c_{\rm V}t)^2}
\end{equation}
and
\begin{eqnarray}\label{compressor_transformation}
\bar{r}=\left\{
                \begin{array}{ll}
                  \frac{r_1}{r_2}r, & 0\le r \le r_2\\
                  ar-b, & r_2 < r \le r_3\\
		     r, & r > r_3
                \end{array}
              \right.
\end{eqnarray}
where $a = (r_3-r_1)/(r_3-r_2)$ and $b = (r_2-r_1)/(r_3-r_2)r_3$. Basically, the circle $r=r_2$ is squeezed into the circle $r=r_1$. To guarantee the continuity of the transformation, the annular region $r_2 < r < r_3$ is expanded to the region $r_1 < r < r_3$ (see Fig.~\ref{Fig2}). The transformation is similar to the one that was employed to compress a region of two-dimensional space~\cite{RHA08-PN}, but it is now applied to compress a region of two-dimensional spacetime (time and one space variable), which significantly changes its meaning. 
Here we will limit the analysis to the propagation of rays, although it could be extended easily to the propagation of waves.

Following the ATA method described in section~\ref{ATA}, we can obtain the relation between the parameters of real and virtual media for a general transformation \mbox{$\left(\bar{t}=f_1(x,t);\, \bar{x}=f_2(x,t)\right)$} as~\cite{GAR13-SR}
\begin{eqnarray}
v_{\rm R}^x&=&\frac{\partial_t{f}_1\partial_t{f}_2-c_{\rm V}^2\partial_x{f}_1\partial_x{f}_2}  {(\partial_t{f}_1)^2-c_{\rm V}^2(\partial_x{f}_1)^2}\bigg|_{\bar{x},\bar{t}\rightarrow x,t},\\
c_{\rm R}^2&=&(v_{\rm R}^x)^2 + \frac{c_{\rm V}^2(\partial_x{f}_2)^2-(\partial_t{f}_2)^2} {(\partial_t{f}_1)^2-c_{\rm V}^2(\partial_x{f}_1)^2}\bigg|_{\bar{x},\bar{t}\rightarrow x,t},\\
\rho_{\rm R}&=&\rho_{\rm V}\frac{c_{\rm R}^2}{c_{\rm V}^2} \frac {(\partial_t{f}_1)^2-c_{\rm V}^2(\partial_x{f}_1)^2} {\partial_t{f}_1\partial_x{f}_2-\partial_x{f}_1\partial_t{f}_2}\bigg|_{\bar{x},\bar{t}\rightarrow x,t}.
\end{eqnarray}
where ${\bar{x},\bar{t}\rightarrow x,t}$ means relabeling $\bar{x},\bar{t}$ to $x,t$. Particularizing these relations to the desired compressing transformation we are led to the sought parameters for the region $r_1 < r < r_3$ (note that $\rho_{\rm R}$ is not relevant in ray acoustics)
\begin{eqnarray}
v_{\rm R}^x&= c_{\rm V}\frac{c_{\rm V}b^2xt(c_{\rm V}^2t^2-x^2)}{r^6+2bc^2_{\rm V} t^2 r^3 + (bc_{\rm V}t)^2(c_{\rm V}^2 t^2-x^2)},\\
c_{\rm R}&=c_{\rm V}\frac{r^3(r^3+br^2)} {r^6+2bc^2_{\rm V} t^2 r^3 + (bc_{\rm V}t)^2(c_{\rm V}^2 t^2-x^2)}.
\end{eqnarray}
Interestingly, for the inner disk ($r < r_1$), we find that $c_{\rm R} = c_{\rm V}$ and $\mathbf{v}_{\rm R} = \mathbf{v}_{\rm V} = \mathbf{0}$, i.e., we do not need to change the background medium in this region. This is in contrast with the devices that implement a uniform compression of space, whose refractive index changes in proportion to the compression factor~\cite{RHA08-PN,GAR11-OE}. The reason is that, in a uniformly compressed spatial region, rays have to travel a shorter distance in the same time, and thus the propagation speed must decrease (the refractive index increases). However, in the case studied here time is also compressed, and rays have to travel this distance in a shorter time as well. Another way to see it is by noticing that, according to the transformation, trajectories have the same slope in the regions $r < r_1$ and $r > r_3$, which implies that they propagate at the same speed (same refractive index). This could be interesting for applications where we want to have the compressed wave directly in the background medium (in this case in the region $r < r_1$, since the compressor is limited to the ring $r_1 < r < r_3$). 

Another remarkable feature of this compressor is that the frequency of the acoustic wave is increased within the disk $r < r_1$ by the same factor as the wavelength decreases (an additional way to explain why the refractive index is not modified there), since time and space are equally compressed. Therefore, we also achieve a frequency shift in an area filled with the background medium, something that does not occur in other transformation-based frequency converters in which only time is transformed~\cite{CUM11-JOPT,GAR14}.

\begin{figure}[]
\begin{center}
  \includegraphics{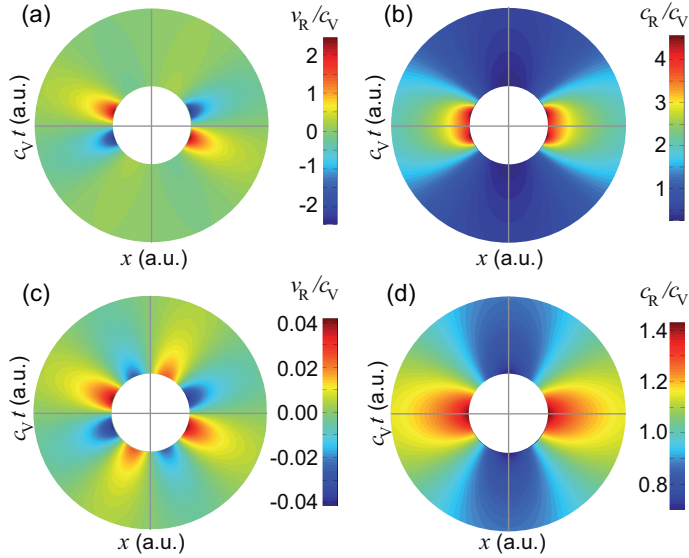}
\end{center}
  \caption{Distribution of the background velocity and speed of sound required for the implementation of a space-time compressor with $r_1 = 0.2$ and $r_3 = 0.6$ for two different values of $r_2$. (a-b) $r_2 = 0.4$. (c-d) $r_2 = 0.25$.}
  \label{Fig3}
\end{figure}

The distribution of background velocity and speed of sound required for the implementation of the space-time compressor is shown in Fig.~\ref{Fig3} for two different values of $r_2$. From the technological point of view, both parameters take a set of reasonable values at any given instant (note that the parameters get more relaxed as the compression factor decreases, i.e., $r_2$ approaches $r_1$). However, the real challenge comes from the fact that the medium properties must change in time. Recent works have demonstrated different ways of achieving a dynamic control of the speed of sound that could be employed for the implementation of ATA devices~\cite{SEI12-APL,CAL14-PNAS}. It would just remain to solve the problem of attaining the desired distribution of background velocities.  In the optical case, a moving medium can be mimicked by using background waves able to modify dynamically the electromagnetic properties of a certain material~\cite{PHI08-SCI}. In principle, a similar strategy could be followed in our acoustic problem.

\begin{figure}[]
\begin{center}
  \includegraphics{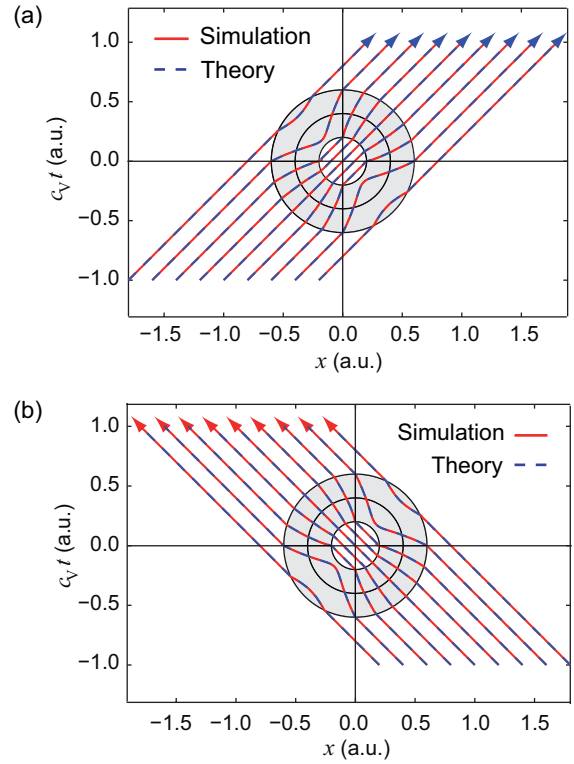}
\end{center}
  \caption{Performance of the proposed spacetime compressor (shaded region). Here, $r_1=0.2$, $r_2=0.4$, and $r_3=0.6$. The device is symmetric and works equally well for rays traveling in (a) the $+x$ direction and (b) the $-x$ direction.}
  \label{Fig4}
\end{figure}

To verify the functionality of the spacetime compressor, we simulated the trajectories followed by different rays going through it (see Fig.~\ref{Fig4}). This was done by solving numerically Hamilton's equations~\cite{GAR13-SR}. The calculated trajectories are compared with the expected theoretical ones (obtained by applying the proposed transformation to the trajectories of the corresponding rays in virtual space, which are just straight lines), finding an excellent agreement. 

Finally, it is worth mentioning that, unlike the time cloaks analyzed in previous studies~\cite{MCC11-JOPT,GAR13-SR}, the proposed compressor works for acoustic waves traveling in both the positive and negative $x$-direction. This is shown in Fig.~\ref{Fig4}. 

\section{Conclusion} \label{Conclusion}

In conclusion, we have seen that, with the help of an auxiliary relativistic spacetime and the tools of analogue gravity, it is possible to build a transformational technique for acoustics as powerful as its electromagnetic counterpart. The main advantage of ATA is that it enables us to design devices based on transformations that mix space and time. As an example, we have shown how this technique can be applied to build a device that increases the density of events within a given spacetime region by simultaneously compressing space and time. Finally, we have discussed some interesting features of this compressor and verified its functionality through numerical simulations.

\section*{Acknowledgement}
This work was developed under the framework of the ARIADNA contract 4000104572/12/NL/KML of the European Space Agency. C.~G.-M., J.~S.-D., and A.~M. also acknowledge support from Consolider project CSD2008-00066, A.~M. from project TEC2011-28664-C02-02, and  C.~B. and G.~J. from the project FIS2011-30145-C03-01. J. S.-D. acknowledges support from the USA Office of Naval Research.

\section*{References}
\bibliography{mybibfile}

\end{document}